\documentclass[journal=pasa,manuscript=research-paper,year=2024,volume=XX,]{cup-journal}
\usepackage{graphics,amsmath,amssymb,multirow,subcaption,hyperref,float,braket}
\usepackage{booktabs}
\usepackage{siunitx}
\hypersetup{colorlinks=true,citecolor=blue,linkcolor=blue,urlcolor=blue}

\title{On the collisional  sensitivity of polarized Mg II solar lines}

\author{M. Derouich}
\author{S. Qutub}
\affiliation{Astronomy and Space Science Department, Faculty of Science, King Abdulaziz University, P.O. Box 80203, Jeddah 21589, Saudi Arabia}
\email[M. Derouich]{aldarwish@kau.edu.sa \& derouichmoncef@gmail.com}


\begin{document}

\begin{abstract}
Neutral and singly ionized states of the Magnesium (Mg) are the origin of several spectral lines that are useful for solar diagnostic purposes. An important element in modeling such solar lines is collisional data of the Mg with different perturbers abundant in the Sun, specially with neutral hydrogen. This work aims at 
providing complete depolarization and polarization and population transfer data for Mg II due to collisions with hydrogen atoms.
For this purpose, a general formalism  is employed to calculate the needed rates of MgII due to collisions with hydrogen atoms. 
The resulting collisional rates are then employed to investigate the impact of collisions on the polarization of 25 Mg II lines
relevant to solar applications
by solving the governing statistical equilibrium equations within multi-level and multi-term atomic models.
We find that the polarization of some Mg II  lines  starts to be sensitive to collisions  for hydrogen density $n_H  \!\gtrsim\!$ 10$^{14}$ cm$^{-3}$. 
\end{abstract}

\section{Introduction }
Magnesium is  abundant in the solar atmosphere, and its neutral and singly ionized states give rise to numerous spectral lines with significant diagnostic potential in the photosphere, chromosphere and the transition region. 
NASA's 
Interface Region Imaging Spectrograph (IRIS) spacecraft provided an important opportunity 
to 
observe intensity spectra 
of several  lines, including the Mg II lines. 
Various works have presented forward modeling of the intensity of Mg II  lines  (e.g., Leenaarts  et al. 2013). 

Polarization spectra of  Mg II h-k lines have been observed  through the Ultra-Violet Spectro-Polarimeter 
on board 
the Solar Maximum Mission  (Calvert et al. 1979, Bohlin et al. 1980,  Woodgate et al. 1980).
These measurements have been analyzed  by Henze \& Stenflo (1987). 
Furthermore,  several theoretical studies have improved our comprehension of the physical processes that cause the polarization of the Mg II h-k lines, highlighting their magnetic sensitivity (e.g. Auer et al. 1980, Henze \& Stenflo 1987, Belluzzi  \& Trujillo Bueno 2012; 
Alsina Ballester et al. 2016; del Pino Alem\'an et al. 2016, 2020).
In addition, 
UV spectropolarimeter called Chromospheric LAyer SpectroPolarimeter (CLASP2) was launched on  2019
to observe polarized light emitted by Mg II ions around the wavelength of 280 nm 
for investigating the magnetic properties of the transition region and the upper chromosphere.

Our aim is to contribute 
to the efforts concentrated on observing and interpreting  Mg II lines by elucidating 
the possible collisional depolarizing role during their formation.  
In fact, 
collisions with hydrogen might be important for modeling chromospheric Mg II lines 
and 
their depolarizing effect has not been elucidated in details previously.

We model 
the Mg II by a  multi-level and then by a  more realistic multi-term  atomic model (see Landi Degl'Innocenti  \& Landolfi 2004).
We provide all needed collisional rates due to the Mg II+H collisions. 
We include these rates in the statistical equilibrium equations (SEE) for  multi-level and  multi-term  atomic models, 
which are then solved to determine how collisions impact the polarization of the Mg II lines.

\section{Collisional effects in multi-level case}
\subsection{SEE collisional contribution}
 We adopt the LS coupling scheme, where the level is usually  denoted by $nl \, ^{2S+1}L_J$, where $n$ is the principal quantum number, $l$ is the orbital angular momentum quantum number, $S$ is the total spin, $L$ is the total orbital angular momentum, and $J$ is the total angular momentum defined as $J = L + S$ (e.g. Martin  \& Wiese  2002). 
For simplicity, we will also use the notations $nl \, ^{2S+1}L_J = nl \, LS\,J = \alpha J$, where $\alpha$ represents the set of quantum numbers $(n \, l \, L S)$. In the context of the polarization studies, the 
representation of Mg II states using the atomic density matrix formalism based on irreducible tensorial operators is demonstrated to be the most appropriate  (e.g. Sahal-Br\'echot 1977; Landi Degl'Innocenti  \& Landolfi 2004).
 In this basis, the contribution of depolarizing isotropic collisions to the SEE is given by (e.g.  Sahal-Br\'echot et al. 2007): 
\begin{eqnarray} \label{eq_ch3_17}
\left( \frac{d \; \rho_q^{k} (\alpha \; J)}{dt}   \right)_{\rm coll} 
\!\!\!\!\!\!&=&\!\!\!\!\!
- D^k(\alpha \; J, T) \; \rho_q^{k} (\alpha \; J) \nonumber \\
&&\!\!\!\!\!
- \rho_q^{k} (\alpha \; J) \sum_{J' \ne J}  \sqrt{\frac{2J'+1}{2J+1}} D^0 (\alpha \; J \to \alpha \; J', T)  \nonumber  \\
&&\!\!\!\!\!
+ \sum_{J' \ne J}  
D^k(\alpha \; J' \to \alpha \; J, T) \;  \rho_q^{k} (\alpha \; J') \, ,
\end{eqnarray}
where $k$, $0 \!\le\! k \!\le\! k_{\rm max}$, is the tensorial order with $k_{\rm max}\!=\! 2J$ for $J\!=\!J'$ and $k_{\rm max}\!=\! min \{2J,2J'\}$  
for  $J \!\ne\! J'$.
$D^k(\alpha \; J, T)$ and $D^k(\alpha \; J \!\to\! \alpha \; J', T)$  are  the depolarization and the polarization transfer rates, respectively.  
Note that $D^0 (\alpha \; J \!\to\! \alpha \; J', T)$ is the population transfer rate between the levels  $(\alpha \; J)$ and $(\alpha \;J')$. 
These 
rates  should be  calculated independently to enter  the SEE.  
To model the Mg II ions, 
we  consider a comprehensive  atomic model containing large number of lines ranging from the ultraviolet to the infrared  as  shown in Figure~\ref{fig1_model_MgII} (see also Fig. 1 of Leenaarts  et al. 2013).

\subsection{Collisional rates for P-, D- and F-states}
\begin{figure}[h]  
\begin{center}
\includegraphics[width=9 cm]{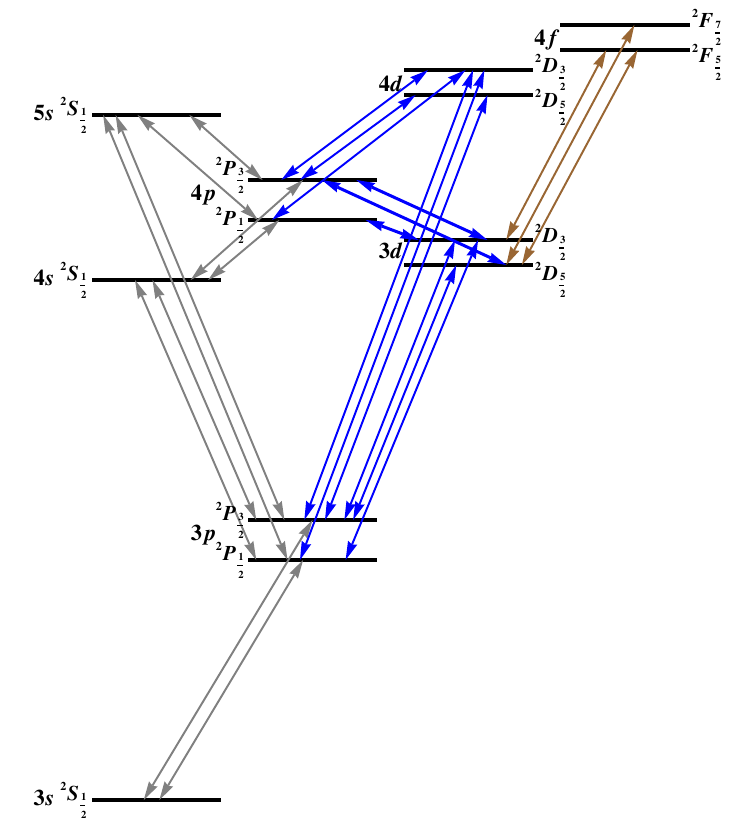}  
\caption{Energy diagram of the  atomic model of Mg II adopted for this work. Note that $p-d$  transitions taken into account are represented in blue, $s-p$ transitions are represented in grey, and $f-d$ transitions are represented in brown color. }
\label{fig1_model_MgII}
\end{center}
\end{figure}

The 
atomic model adopted in this paper contains $s$-, $p$-, $d$- and $f$-states (see Figure~\ref{fig1_model_MgII}). 
Note that $s$-states of the present model are not linearly polarizable and, therefore, are not affected 
by collisions with hydrogen atoms.
The rates $D^k(\alpha \; J, T)$ and $D^k(\alpha \; J' \!\to\! \alpha \; J, T)$ for the levels of the $3p \; ^2P$ term 
have been previously calculated using the hybrid method developed by Derouich (2020). 
In order to obtain the rates associated to the other terms of Figure~\ref{fig1_model_MgII}:   
$4 p \; ^2P$,  $3 d \; ^2D$, $4 d \; ^2D$ and $4 f \; ^2F$, 
one can either 
employ a direct calculation by running the numerical code of collisions 
for each level as explained in the formalism presented by Derouich et al. (2004) 
or use the genetic programming (GP) functions presented in Derouich (2017) for $p$-states, 
Derouich et al. (2017) for $d$-states and Derouich (2018) for $f$-states.

Whether running the numerical  code of collisions   or using the GP functions, it is necessary to determine the Uns\"old energy $E_P$  for each ionic state. This energy must then be used as input for either the numerical code or the GP functions. In addition to $E_P$, the effective quantum number $n^*$ is also needed as input parameter.  
 If the energy of the state $\ket{a}$ of the valence electron is $E_{a}$ and the ionization energy of the ion is $E_\infty$, the effective quantum number $n^*$  is given by (see, e.g., Derouich 2004):
\begin{equation}  \label{eq_ch3.1}
n^*= 2 \times [2 (E_\infty - E_{a })]^{-1/2} 
\end{equation}
where the energies are in atomic units.
Values of $n^*$ are provided in Table~\ref{table_Ep_n} for all the cases where $n^*$ is required to determine the collisional rates. Appropriate value of $E_p$ can then be calculated from the Uns\"old formula (see, e.g., Barklem \& O'Mara 1998):
\begin{equation} \label{eq_ch6_6}
E_p = - \frac{2 <\! r^2 \!>}{C_6},
\end{equation}
in atomic units, where  $C_6$ is the van der Waals coefficient  (see Table~\ref{table_Ep_n}), and $<\! r^2 \!>$ is the average of the squared distance between the electron of valence and the nucleus of the perturbed ion (see Table~\ref{table_Ep_n}). 
By adopting the hydrogenic approximation,  one has for singly ionized atoms (see, e.g., Barklem   1998):
\begin{equation} \label{eq25}
<\! r^2 \!>= \frac{n^{*2}}{8} [5n^{*2}+1-3l(l+1)].
\end{equation}
The van der Waals coefficients  are  available on Kurucz's website 
(Kurucz 2013; \href{http://kurucz.harvard.edu/atoms/1201/}{kurucz.harvard.edu}), namely the file ``\href{http://kurucz.harvard.edu/atoms/1201/gammasum1201z.gam}{gammasum1201z.gam}'' which contains comprehensive information on the Mg II lines and relevant atomic parameters.
For the $3 d$  $^2D$ state, 
\begin{equation} \label{eq25}
C^{\rm Kurucz}_6(\text{ Mg II},  3 d  \;\;  ^2D) = 1.93  \times  10^{-32} \; \text{cm}^6 \; \text{s}^{-1}
\end{equation}
Note that the definition of $C_6$ included in  Equation~\ref{eq_ch6_6} giving $E_p$ differs by a factor of $h$ ($h$ is the Planck constant) from  $C^{\textrm{Kurucz}}_6$, in addition to the difference in units since $C_6$ of Equation~\ref{eq_ch6_6} is in a.u. while $C^{\textrm{Kurucz}}_6$ is in $\text{cm}^6 \; \text{s}^{-1}$. In fact,
\begin{eqnarray} \label{eq25} 
C_6 (\text{ Mg II},  3 d   \;\;  ^2D) 
\!\!\!\!&=&\!\!\!\!
6.92 \times 10^{33} \;  C^{\textrm{Kurucz}}_6(\text{ Mg II},  3 d  \;\;  ^2D) \nonumber \\ 
\!\!\!\!&=&\!\!\!\! 133.56 \; \text{ a.u.},  
\end{eqnarray}
implying that:
\begin{eqnarray} \label{eq25} 
E_p = -0.464  \text{ a.u.}  
\end{eqnarray}
Similar calculation allows us to obtain all $E_p$ values needed for the computation of depolarization and polarization and population transfer rates (see Table~\ref{table_Ep_n}).  

 Our depolarization and polarization transfer rates are written in the form 
 $D^k=a^k \times 10^{-9}  \; n_H   \left(\! \frac{T}{5000} \!\right)^{\lambda^k}$. Note that $a^k \times 10^{-9}=D^k/n_H$ at $T=5000$ K and $\lambda^k$ is   the so-called velocity
exponent (e.g. Derouich et al. 2004). Values of  $a^k$ and $\lambda^k$   with even $k$-order, 
which are relevant for linear polarization treatment within the multi-level model, 
are provided in Table~\ref{table_multilevel} (see \ref{coll_data}).   
In particular, 
we provide only $D^k(\alpha \;J \!\to\! \alpha \;J')$ since $D^k(\alpha \;J' \!\to\! \alpha \;J)$ are calculated by applying  the detailed balance relation:
\begin{equation}  \label{eq_11}
D^k(\alpha \;J' \!\to\! \alpha \;J) =   \frac{2J+1}{2J'+1}  \exp \left(\frac{E_{J'}-E_{J}}{k_BT}\right) D^k(\alpha \;J \!\to\! \alpha \;J')
\end{equation}
with  $k_B$ is the Boltzmann 
constant and $E_{J}$ is the energy of the $J$-level.
\begin{figure}[h]
\begin{center}
\includegraphics[width=8.5 cm]{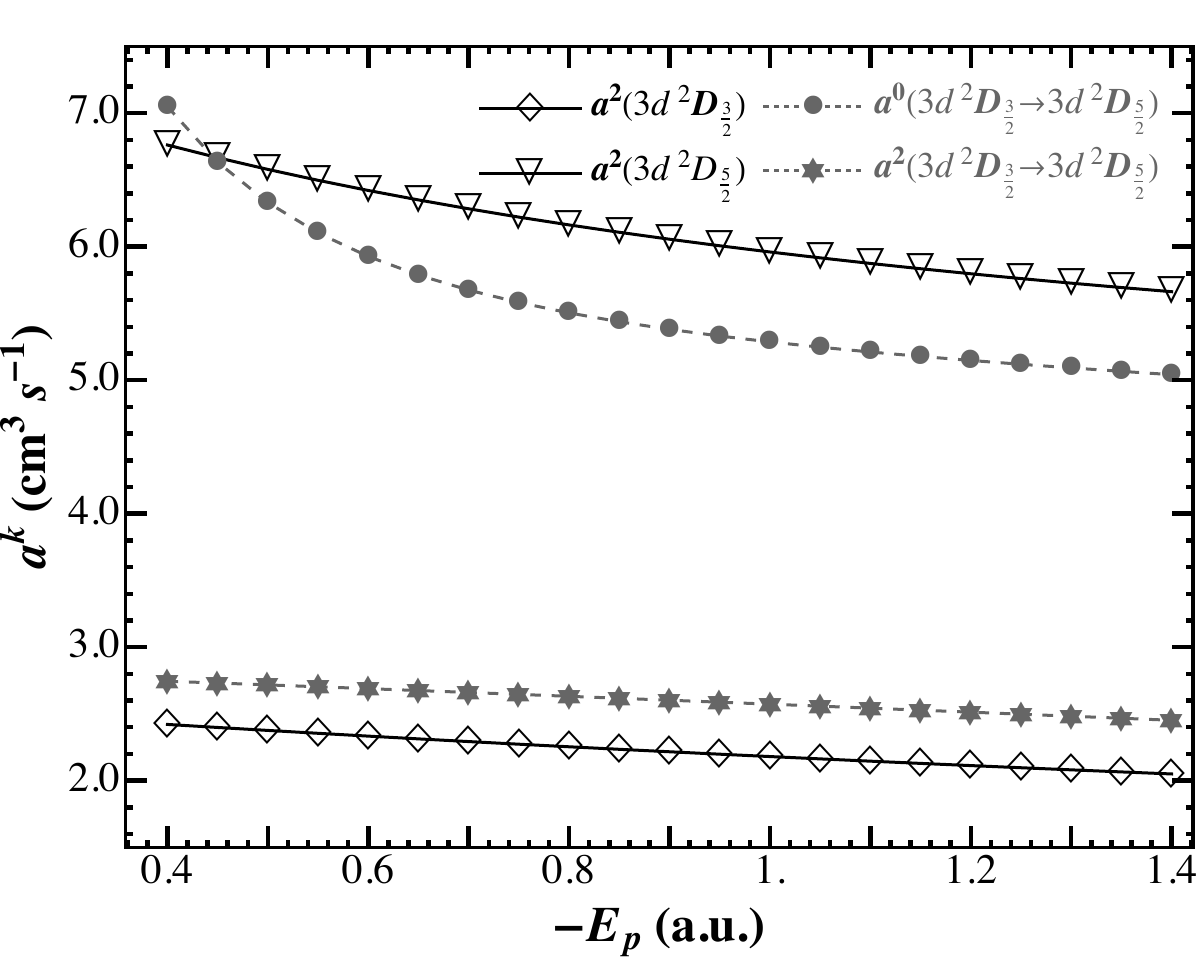}  
\caption{Variation of the collisional rates with $E_p$.}
\label{Fig-1-Ep}  
\end{center}
\end{figure}
\begin{table}
\begin{center}
\begin{tabular}{l c  c c r}
\hline
\hline
Term &$E_p$ (a.u.) & $C_6$ (a.u.)  &$n^*$ & $<\! r^2 \!> $   \\
 &  &    & &   (a.u.) \\
\hline  
 $4p$ $^2P$   &  -0.42665  & 310.02     &  3.286 &   66.12  	 \\
 $3d$ $^2D$   & -0.4636   &   133.56    &   2.990  &30.96	 \\
 $4d$ $^2D$   &   -0.4496  &     536.99   & 3.962    & 120.71 	 \\
 $4f$ $^2F$   &   -0.4431   &   404.82     &  3.997   & 89.69	  	 \\
\hline  
\end{tabular}
\end{center}
\caption{\textsf{Mg II atomic and collisional parameters for the multi-level case}}
\label{table_Ep_n}
\end{table} 

Note that 
the collisional rates as they are presented in this section depend on the precision of the calculation of $E_p$,
which
is related to the precision of $C_6$ obtained from the Kurucz methodology and $<\! r^2 \!>$ resulting from the hydrogenic approximation.  
To 
examine the dependence of the 
depolarization rates \\ $D^{2}(3d \; ^2D_{\frac{3}{2}})$ 
and $D^{2}(3d \; ^2D_{\frac{5}{2}}) $
and
  transfer rates 
$D^{2}(3d \; ^2D_{\frac{3}{2}} \!\to\!  3d \; ^2D_{\frac{5}{2}})$
and 
$D^{0}(3d \; ^2D_{\frac{3}{2}} \!\to\!  3d \; ^2D_{\frac{5}{2}}) $
to $E_p$, we study the sensitivity of the $a^k$ coefficients to the $E_p$ variation. 
As it can be seen in Figure~\ref{Fig-1-Ep}, 
the $a^k$  are sufficiently stable with respect to sensible variation of $E_p$.  
The range of the calculation presented in Figure~\ref{Fig-1-Ep} is chosen to contain possible values of $E_p$, 
based on different models, 
such as the one adopted by O'Mara \& Barklem (1998) for the
$3d \; ^2D$ 
state of the Ca II, where $E_p$ was found to be between -0.918 and -1.236 a.u.
\begin{figure}[h]
\begin{center}
\includegraphics[width=9 cm]{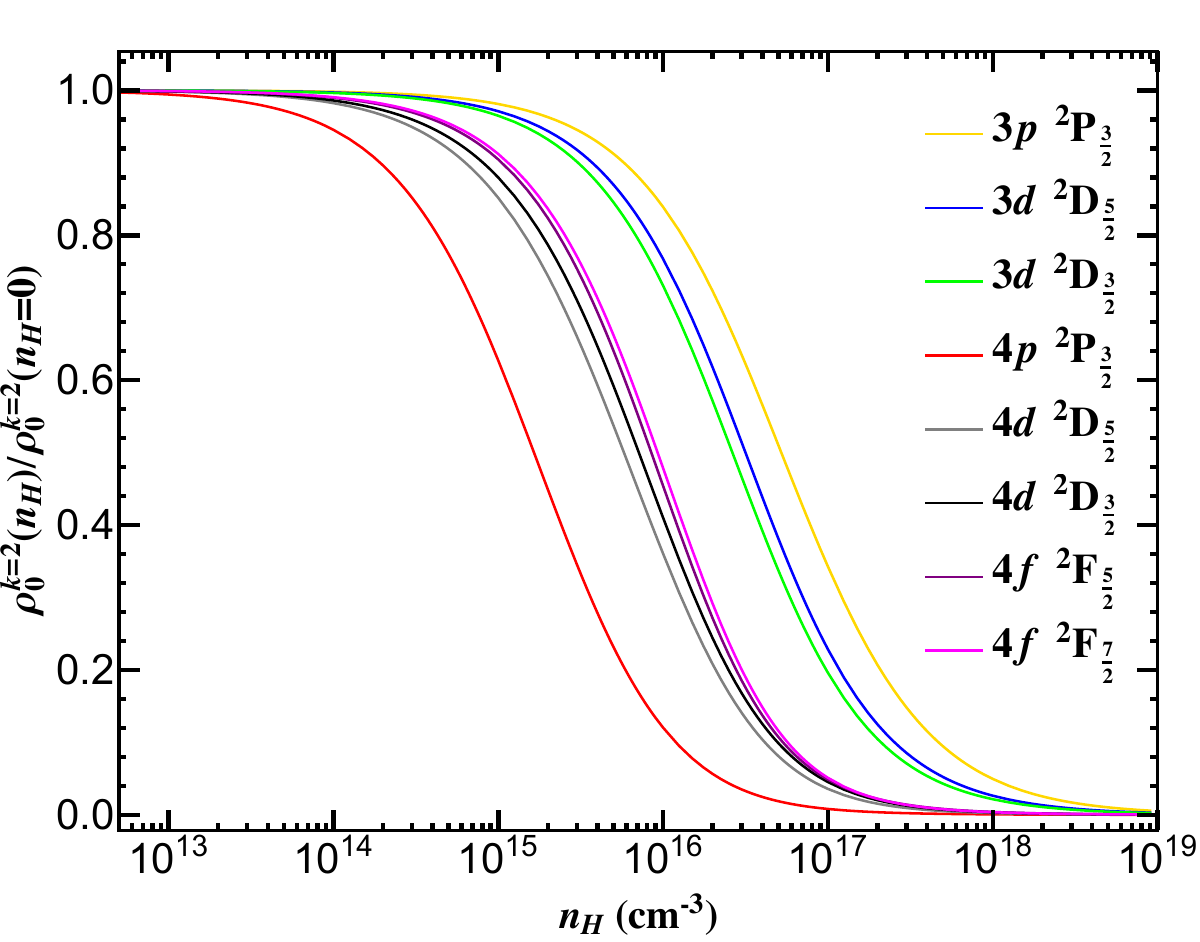}  
\caption{Variation of the ratio $\frac{\rho_0^{2}(n_H)}{\rho_0^{2}(n_H=0)} $  as a function  of $n_H$ for the Mg II levels within the framework of the multi-level model.}
\label{Fig-MgII-paper_1}  
\end{center}
\end{figure}
%
%
%

\subsection{Resolution of the SEE in the multi-level case}
We 
consider a slab of Mg II ions positioned at the solar atmosphere, and we assume a zero-magnetic field case.  
To 
generate a polarized light, the slab of Mg II  ions is assumed to be illuminated from below by anisotropic solar radiation.
The components of the incoming radiation field are typically represented by $J_{q}^{k} $,   where $k$ is the tensorial order and $q$ signifies the coherences in the tensorial basis ($-k \!\le\! q \!\le\! k$).   
 The atmospheric light is considered to be uniform and
 exhibits cylindrical symmetry about the local solar vertical at the scattering center (the Mg II ion)
which  
implies that, for a given frequency $\nu$, only $J^0_0(\nu)$ and $J^2_0(\nu)$  are non-zero.  
The components
$J^0_0(\nu)$ and $J^2_0(\nu)$  are given by   the number of photons per mode
$\bar{n}(\nu) = J^0_0 \; (c^2/2hJ^0_0(\nu)^3)$ 
and  
the value of the anisotropy factor  
$w(\nu) = \sqrt{2} \; (J^2_0/J^0_0)$   (see e.g. Derouich et al. 2007). 
Calculations of the non-zero components of the incident radiation field,  $J^0_0(\nu)$ and $J^2_0(\nu)$, for each line are based on  Cox (2000) (see also   Allen \& Cox  1999, Thuillier et al. 2004).

The anisotropic incident radiation field induces population imbalances and coherences between the Zeeman sublevels of a given $(\alpha J)$-level of the Mg II ions. These population imbalances and coherences are referred to as atomic polarization. Under these conditions, owing to the cylindrical symmetry of the problem, only linear polarization is produced by scattering, and only the density matrix elements with  with $q=0$ and $k$ even are non-zero.
The statistical equilibrium equations (SEE) include 26 unknowns, corresponding to the density matrix elements $ \rho_0^{k}$.

The Einstein coefficients required for the radiative rates entering the SEE are extracted from NIST database. In addition  
to the expressions of the radiative rates, which are taken from Landi Degl'nnocenti \& Landolfi (2004), the numerical code incorporates the collisional rates, given in Table~\ref{table_multilevel},
as input to compute the values of the atomic density matrix elements. 
These values 
are 
affected by the gain terms called collisional polarization transfer rates and denoted by $D^k(\alpha \; J' \!\to\! \alpha \; J)$, and  by   the loss (collisional relaxation) terms  $D^k(\alpha \; J)+\sqrt{\frac{2J'+1}{2J+1}} D^0 (\alpha \; J \!\to\! \alpha \;  J', T)$ (see Equation~\ref{eq_ch3_17}). 
 
 Our focus is on the effects of elastic collisions with neutral hydrogen to complement existing studies that omit these collisions. While a more comprehensive analysis of Mg II line polarization would include processes such as  collisions with electrons and polarized radiative transfer in a magnetized atmosphere; these are beyond the scope of this paper.  
 
 Given that collisional rates are proportional to the hydrogen density $n_H$ (according to the impact approximation), analyzing how polarization depends on collisional rates is equivalent to examining its dependence on $n_H$. 
In theory, 
even if the core of  a line  is formed  at higher chromospheric layers,   
its wings can be formed in deeper layers of the chromosphere. 
Given that 
the density of hydrogen atoms $n_H$  is larger in deep chromospheric layers and 
that the considered lines cover spectral window from the ultraviolet to the infrared, one should perform a scan over large range of $n_H$ values. 
We 
solve the SEE to calculate the non-zero $ \rho_0^{k}$ elements by adopting the multi-level atomic model  
presented in the Figure~\ref{fig1_model_MgII} for a wide range of hydrogen density values $n_H$ going from $10^{12}$ cm$^{-3}$ to  $10^{19}$ cm$^{-3}$ while adopting a solar temperature  $T=$ 5000 K. 
We point out the range of $n_H$ where the effect of collisions is important.


In Figure~\ref{Fig-MgII-paper_1}, we shows the ratio $\frac{\rho_0^{k=2}(n_H)}{\rho_0^{k=2}(n_H=0)}$ giving the variation of the alignment as a function of $n_H$. 
It can be   seen that collisions start 
 influencing the alignment of the $4p \; ^2P_{3/2}$ level for  $n_H$ $\sim$ $10^{14}$ cm$^{-3}$. Consequently, the 6 lines directly  connected to this level (see Figure~\ref{fig1_model_MgII}) should be affected by collisions for  $n_H$ $\gtrsim$ $10^{14}$ cm$^{-3}$.
Lines with upper and/or lower levels different from the $4p$  $^2P_{3/2}$ could be also indirectly slightly affected for  $n_H$ $\gtrsim$ $10^{14}$ cm$^{-3}$ due to the coupling between the different levels through the SEE. All the levels other than $4p \; ^2P_{3/2}$ start to be depolarized  for  $n_H$ $\gtrsim$ $10^{15}$ cm$^{-3}$.

\section{Multi-term case}
In this section, we will allow for coherences between different $(\alpha J)$ and $(\alpha J')$-levels   grouped within the same term  $n\, l \, ^{2S+1}L$. The density matrix elements associated to these coherences are  denoted by $\rho_q^{k} (\alpha \; JJ')$. This is the so-called  mutli-term case (see Section 7.5 of Landi Degl'Innocenti  \& Landolfi  2004).   Indeed, a more 
realistic diagnostic of the polarization of the lines of Mg II should be performed in the framework of the multi-term atomic model due to the tiny energy separation between the levels of the $d$- terms and similarly between those of the $f$- terms.
We   
consider the atomic model of Mg II shown in  Figure~\ref{fig1_model_MgII}, with eight terms: $3 s$  $^2S$, $3 p$  $^2P$, $4 s$   $^2S$, $3 d$  $^2D$, $4 p$  $^2P$, $5 s$  $^2S$,  $3 d$  $^2D$ and $4 f$  $^2F$; 
we take into account coherences between different $J$-levels within these terms. 
In fact, 
In 
the spherical statistical tensors representation, the contribution of  the isotropic collisions to the SEE in the multi-term case is:
\begin{eqnarray} \label{eq_LL}
\left(\! \frac{d \; \rho_q^{k} (\alpha \; JJ')}{dt} \!\right)_{\!\rm coll} 
\!\!\!\!\!\!& = &\!\!\!\!\!  
- \Big[\!\! \sum_{(J''J''')\ne (JJ')} \!\! \sqrt{ \frac{J''+J'''+1} {J+J'+1}} \; D^0(\alpha \; JJ' \!\to\! \alpha \; J''J''')  \nonumber \\ 
&&\!\!\!\!\!
+ D^k(\alpha \; JJ') \Big] \times \rho_q^{k} (\alpha \; JJ')  \\
&&\!\!\!\!\!
+\! \sum_{(J''J''')\ne (JJ')} \!\!\!\! D^k(\alpha \; J''J''' \!\to\!  \alpha \; JJ')   \times \rho_q^{k} (\alpha \; J''J''')   \nonumber
\end{eqnarray}
Note that $J, J', J''$, and $J'''$ represent possible values of the total angular momentum  within the same term  $n\, l \, ^{2S+1}L$.

The radiative contributions are taken from multi-term atomic model as described by Landi Degl'Innocenti  \& Landolfi (2004).  
In order to 
calculate $\rho_q^{k} (\alpha \, JJ')$,
we need  the depolarization rates $D^k(\alpha \,JJ')$ 
and the polarization/population transfer rates $D^k(\alpha \,JJ' \!\to\! \alpha \, J''J''')$  with $(JJ') \ne (J''J''')$ due to collisions of  H atoms
in their ground state $^2S_{1/2}$ 
with Mg II presented by the eight-term atomic model adopted in this work. 
Direct calculation of the 
$D^k(\alpha \, JJ')$ and  $D^k(\alpha \, JJ' \!\to\! \alpha \, J''J''')$ 
rates is a complicated task since 
one should take into account the coherences between $J$-levels  
when calculating the interaction potential and solving the Schr\"odinger equation. 
At the best of our knowledge, this direct calculation have not been performed neither theoretically nor experimentally. 
To address this problem, 
one can adopt an  indirect and more practical method, based on the frozen spin $S$ approximation. 
Using this approach, one can show that:
\begin{eqnarray}\label{eq_frozen}
D^{k} (\alpha \, JJ' \!\to\!  \alpha \, J''J''') \!\!\!\!\!&=&\!\!\!\!\! \sqrt{(2J + 1) (2J' + 1) (2J'' + 1) (2J''' + 1)}  \nonumber  \\ 
&&\!\!\!\!\!\!\!\!\!\!\!\!\!\!\!\! \times
\sum_{k_L} (2k_L + 1) \, D^{k_L} (n \, l \, L)   \\
&&\!\!\!\!\!\!\!\!\!\!\!\!\!\!\!\!  \times
\sum_{k_S} 
(2k_S + 1) \left \{\!\!
\begin{array} {ccc}
  L & S & J \\
 L & S & J' \\
 k_L & k_S & k
\end{array}
\!\!\right\}
\!\!
\left \{\!\!
\begin{array} {ccc}
  L & S & J'' \\
 L & S & J''' \\
 k_L & k_S & k
\end{array}
\!\!\right\} \nonumber
\end{eqnarray}
Equation~\ref{eq_frozen} is obtained  through a methodology which is formally similar to that firstly proposed by  Nienhuis  (1976) and Omont (1977) (see also  Derouich  \& Barklem 2007 and Sahal-Br\'echot et al.  2007). 
We note that diagonal rates $J=J'$ and  $J''=J'''$  are similar to the transfer rates used for the case of multi-level atomic models. 
However, 
calculation of the off-diagonal rates  (i.e., $J \ne J'$ and/or  $J''  \ne  J'''$) requires the evaluation of the  tensorial rates $D^{k_L} (n\, l \,L)$ (see Equation~\ref{eq_frozen}). Since the Mg II is a simple ion  modelled as a single valence electron outside a spherical ionic core, one has  $D^{k_L} (n\, l \,L)=D^{k_l} (n\, l)$ where $L=l$ is the angular momentum of the optical electron (see Derouich et al. 2005 and Sahal-Br\'echot et al. 2007).

The main step to obtain  $D^{k_l} (n\, l)$, needed in Equation~\ref{eq_frozen}, is the calculation of the scattering matrix. We ran our numerical code to solve the coupled differential equations and obtain the scattering matrix elements in the basis  $\ket{l m_{l}}$  for a given velocity and impact parameter $b$ (where $ m_{l}$ is the projection of $l$ along the quantization axis; see Derouich et al. 2003a,b; Derouich et al. 2004). 
We then obtained the $D^{k_l} (n\, l)$ rates through an integration over the impact parameter $b$ and the Maxwell distribution of velocities (see Equations 7, 9, 12, and 20 of Derouich et al. 2003a).

\begin{figure}[h]
\begin{center}
\includegraphics[width=9 cm]{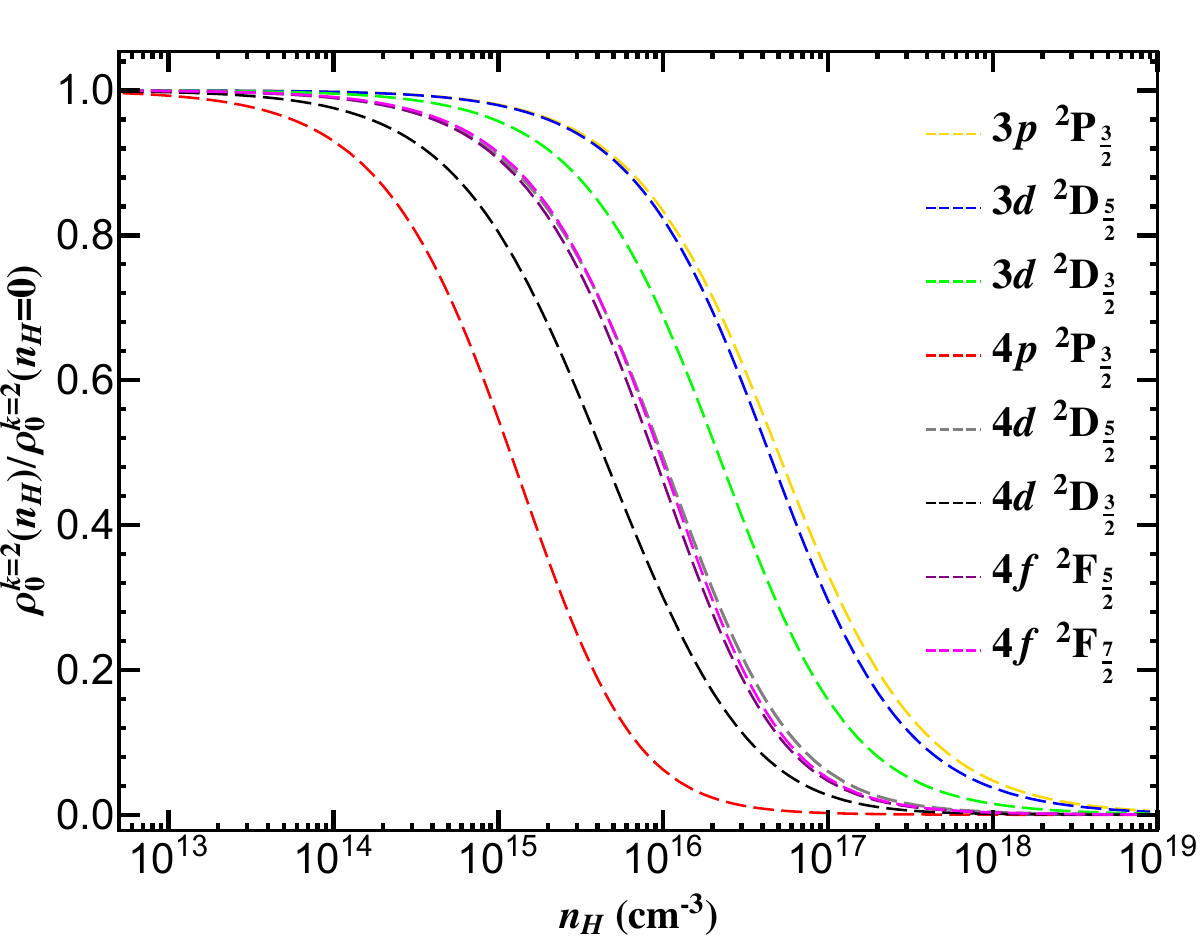}  
\caption{Same as Figure~\ref{Fig-MgII-paper_1} but for the multi-term model.}
\label{Fig-MgII-paper_2}  
\end{center}
\end{figure}

\begin{figure}[h]
\begin{center}
\includegraphics[width=9 cm]{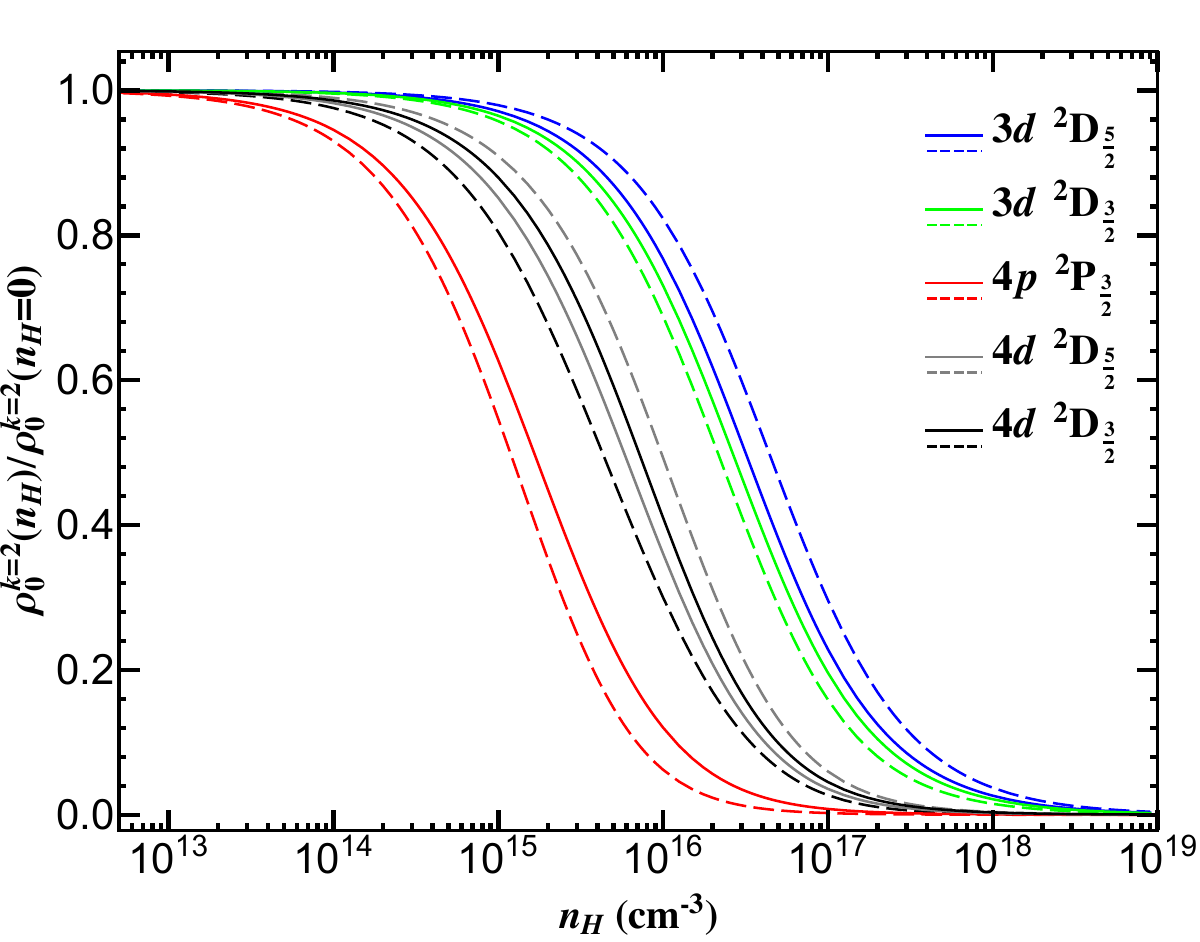}  
\caption{Comparison of the ratio $\frac{\rho_0^{2}(n_H)}{\rho_0^{2}(n_H=0)}$ as a function of $n_H$ for the multi-term and multi-level cases. The dashed line represents the multi-term case, while the solid line represents the multi-level case.}
\label{Fig-MgII-paper_3}  
\end{center}
\end{figure}

All the non-zero depolarization and polarization transfer rates are written in the form 
\begin{eqnarray} \label{eq_MT}
D^k=a^k \times 10^{-9}  \; n_H   \left(\! \frac{T}{5000} \!\right)^{\lambda^k}
\end{eqnarray}
and are tabulated in the Tables~\ref{table_3p_4p}, \ref{table_3d}, \ref{table_4d}, and \ref{table_4f}.
Note that, in particular, $ D^{k} (\alpha \, JJ  \!\to\!   \alpha \, J'J')= D^{k} (\alpha \, J  \!\to\! \alpha \, J')$ are related to  $ D^{k} (\alpha \, J'J'  \!\to\! \alpha \, JJ) $ through the usual detailed balance relation (see Equation~\ref{eq_11}).
We compute all the non-zero  density matrix elements with even $k$-order which are 44 in the case of our multi-term atomic model of Figure~\ref{fig1_model_MgII}. The results obtained for some diagonal elements $\rho_q^{k=2} (\alpha JJ)$  are presented in Figure~\ref{Fig-MgII-paper_2}. 
By 
comparing with Figure~\ref{Fig-MgII-paper_1}, 
one can see that effects of collisions in the multi-level and multi-term cases have the same behaviors. 
However,  
for some cases presented in the Figure~\ref{Fig-MgII-paper_3},
the effect of collisions in the multi-term case is quantitatively different from that in the multi-level case.
In particular, 
the levels  $4 p$  $^2P_{3/2}$, $3 d$  $^2D_{3/2}$, and $4 d$  $^2D_{3/2}$ are more sensitive to collisions in the multi-term case compared to the case of multi-level model. On the other hand, 
the levels  $3d$  $^2D_{5/2}$ and $4d$  $^2D_{5/2}$ are less sensitive to collisions in the multi-term case.

As shown above there are some differences in the sensitivity of Mg II levels to collisions with hydrogen atoms 
between the multi-term and multi-level cases;
nevertheless, 
these differences are practically rather small, specially for sufficiently low $n_H$.
For 
the case of the well known chromospheric Mg II UV h-k lines which are directly connected to the terms $3 s$  $^2S$ and $3 p$  $^2P$, we find that collisions are not important and can be safely neglected in future studies. 
In fact,   
the term $3 s$  $^2S$ is not sensitive  to collisions with hydrogen atoms because of its total angular momentum being $1/2$, meaning the depolarizing  rates $D^{k=2}$ are zero by definition. 
Additionally,  
for the $s$-term there is no collisional transfer within the term since it only has one fine structure level. 
Moreover,
the $3 p$  $^2P$, 
as it is shown in Figures~\ref{Fig-MgII-paper_1} and \ref{Fig-MgII-paper_2}, it starts to be sensitive to collisions only for $n_H \!\gtrsim\! 10^{16}$ cm$^{-3}$, which is higher than the typical densities encountered in the chromosphere. 
%
%
%
%
%
%
%

\section{Conclusion}
The role of the collisional processes for the Mg II lines is usually disregarded when solving the master equation for the atomic density matrix. 
We find that the polarization of Mg II h and k lines are not affected by collisions with neutral hydrogen.
Nevertheless, we find that collisions with hydrogen atoms are important for other Mg II lines, specially those directly connected to the level $4 p$  $^2P_{3/2}$. Our result could help eliminate potential sources of error or uncertainty 
and 
provide clarity on the true significance of collisions for future studies.  
We provide all the collisional rates for the multi-level (13 levels) and multi-term (8 terms) models of Mg II in \ref{coll_data} of this paper.
\section*{Acknowledgements}
This research work was funded by Institutional Fund
Projects under grant no. (IFPIP:995-130-1443). The authors gratefully acknowledge technical and financial support provided by the Ministry of Education and
King Abdulaziz University, DSR, Jeddah, Saudi Arabia


\newpage
\begin{appendix}
\section{Collisional data} \label{coll_data}
\begin{table}[h!]
\begin{center}
\begin{tabular}{l c c c c c c r}
\hline
\hline
Term &$J$ & $J' $ &$k$&  $a^k$ & $\lambda^k$  \\
 \hline 
 $3p$ $^2P$    &  3/2   &  3/2      &  2  & 3.92    &    0.372   	 \\
                        &  1/2   &  3/2      &  0  &   2.94   &   0.493   	 \\
\hline 
$3d$ $^2D$  
                 &  3/2   &  3/2      &  2  &  7.43  &    0.377	 \\
                 &  5/2   &  5/2      &  2  &    6.62  & 0.383   	 \\
                  &  5/2   &  5/2     &  4  &  7.10   &   0.379 		 \\
                                    &  5/2   &  3/2      &  0  &  6.52    &   0.382 	 \\
                &  5/2   &  3/2      &  2  &   2.75    &   0.376 	 \\
                 \hline 
 $4p$ $^2P$    &  3/2   &  3/2      &  2  & 23.26   &  0.377    	 \\
                 &  1/2   &  3/2      &  0  &    16.71   &      0.367 	 \\
                 \hline 
$4d$ $^2D$  
                 &  3/2   &  3/2      &  2  &  14.26  &     0.388	 \\
                 &  5/2   &  5/2      &  2  &    10.75   &    0.382 	 \\
                  &  5/2   &  5/2     &  4  &    11.43 &     0.376		 \\
                                    &  5/2   &  3/2      &  0  & 10.26    &     0.366	 \\
                &  5/2   &  3/2      &  2  &    4.62   &     0.370	 \\
                                 \hline 
$4f$ $^2F$  
                 &  5/2   &  5/2      &  2  &   5.80 &   0.401 	 \\
                 &  5/2   &  5/2      &  4  &    5.55   &   0.373  	 \\
                  &  7/2   &  7/2     &  2  &  9.99   &  0.367   		 \\
     &  7/2   &  7/2     &  4  &   4.65  &     0.390		 \\
                     &  7/2   &  7/2     &  6  &  2.90   &   0.383   		 \\
      &  5/2   &  7/2      &  0  &   12.02  &    0.376 	 \\
       &  5/2   &  7/2      &  2  &    1.34  &     0.379 \\
            &  5/2   &  7/2      &  4  &   0.44  &    0.367  	 \\
  \hline  
\end{tabular}
\end{center}
\caption{\textsf{Mg II collisional rates for the multi-level case.}}
\label{table_multilevel}
\end{table} 
\begin{table}
\begin{center}
\begin{tabular}{l c c c c c c r}
\hline
\hline
Term &$J$ & $J' $ &$J''$ & $J''' $ &$k$&  $a^k$ & $\lambda^k$  \\
 \hline  
 $3p$ $^2P$   &  0.5   &  1.5   & 0.5   &  1.5   &  2  &    3.785  &    0.308  	 \\
      &  1.5   &  0.5   &  1.5   &  0.5    &  2  &   3.785   &     0.308 	 \\
     &   1.5   &  1.5   & 1.5   &  1.5    &  2  &  3.92    &    0.372  	 \\
           &  0.5   &  0.5   & 1.5   &  1.5    &  0  &   2.94    &   0.493  	 \\ 
  &  0.5   &  1.5   & 1.5   &  0.5    &  2  &     -0.045 &     0.210 	 \\
 &  0.5   &  1.5   & 1.5   &  1.5    &  2  &    0.09  &   0.348  	 \\
   &  1.5   &  0.5   & 0.5   &  1.5    &  2  &   -0.045   &   0.210   	 \\
  &  1.5	&0.5  & 1.5&	1.5    &  2  &     -0.09  &      0.348 	 \\
&  1.5	&1.5  & 0.5&	1.5    &  2  &  0.09    &      0.348 	 \\
   &  1.5	&1.5  & 1.5&	0.5    &  2  &   -0.09   &   0.348    	 \\
\hline 
 $4p$ $^2P$   &  0.5   &  1.5   & 0.5   &  1.5   &  2  &    15.45  & 0.337      	 \\
      &  1.5   &  0.5   &  1.5   &  0.5    &  2  &   15.45   &    0.337   	 \\
      &   1.5   &  1.5   & 1.5   &  1.5    &  2  &    23.26   &  0.377     	 \\
           &  0.5   &  0.5   & 1.5   &  1.5    &  0  &   16.71    &      0.367	 \\
  &  0.5   &  1.5   & 1.5   &  0.5    &  2  &     -2.60 &  0.295   	 \\
 &  0.5   &  1.5   & 1.5   &  1.5    &  2  &    5.20  &     0.291	 \\
   &  1.5   &  0.5   & 0.5   &  1.5    &  2  &   -2.60   &   0.295  	 \\
  &  1.5	&0.5  & 1.5&	1.5    &  2  &     -5.20   &     0.291	 \\
&  1.5	&1.5  & 0.5&	1.5    &  2  &  5.20     &   0.291   	 \\
   &  1.5	&1.5  & 1.5&	0.5    &  2  &    -5.20  &    0.291  	 \\
\hline  
\end{tabular}
\end{center}
\caption{\textsf{P-states: Mg II collisional rates for the multi-term case.}}
\label{table_3p_4p}
\end{table} 
\begin{table}
\begin{center}
\begin{tabular}{l c c c c c c r}
\hline
\hline
Term &$J$ & $J' $ &$J''$ & $J''' $ &$k$&  $a^k$ & $\lambda^k$  \\
\hline
 $3d$ $^2D$  &    2.5   &  1.5    & 2.5   &  1.5       &   2  &  8.34   &   0.314 	 \\
&  2.5   &  1.5    & 2.5   &  1.5       &   4  &  10.68  &     0.310	 \\
&  1.5   &  2.5    & 1.5   &  2.5       &   2  &    8.34   &     0.314 	 \\
&  1.5   &  2.5    & 1.5   &  2.5       &  4  &    10.68  &    	0.310         \\
&  1.5   &  1.5    & 1.5   &  1.5       &  2  & 7.43   &    	0.377        \\
&  2.5   &  2.5    & 2.5   &  2.5       &   2  &   6.62  & 0.383   	 \\
&  2.5   &  2.5    & 2.5   &  2.5       &   4  & 7.10   &   0.379 	 \\
&  2.5   &  2.5    & 2.5   &  1.5       &   2  &  2.91    &    	0.367	 \\
&  2.5   &  2.5    & 2.5   &  1.5       &   4  &   1.45   &    	0.299	 \\
&  2.5   &  2.5    & 1.5   &  2.5       &   2  &    -2.91  &    	0.367	 \\
&  2.5   &  2.5    & 1.5   &  2.5       &   4  &  -1.45   &   	0.299 	 \\
&  2.5   &  2.5    & 1.5   &  1.5       &   0  & 6.52    &   0.382 	 \\
&  2.5   &  2.5    & 1.5   &  1.5       &   2  & 2.75    &   0.376 	 \\
&  2.5 & 1.5   & 2.5 & 2.5       &   2  & 2.91     &    0.367	 \\
&  2.5 & 1.5   & 2.5 & 2.5       &   4  &   1.45   &     0.299	 \\
   &  2.5 & 1.5   & 1.5 & 2.5       &   2  &   3.30  &    0.219	 \\
   &  2.5 & 1.5   & 1.5 & 2.5       &   4  &   0.513   &     0.255	 \\
    &  2.5 & 1.5   & 1.5 & 1.5       &   2  &   0.506   &    0.221 	 \\
    &  1.5 & 2.5   & 2.5 & 2.5       &   2  &   -2.91  &  0.367  	 \\
  &  1.5 & 2.5   & 2.5 & 2.5       &   4  & -1.45   &    	0.299 	 \\
     &  1.5 & 2.5   & 2.5 & 1.5       &   2  &    3.30  &   0.219  	 \\
     &  1.5 & 2.5   & 2.5 & 1.5       &   4  &   0.513   &     0.255	 \\
    &  1.5 & 2.5   & 1.5 & 1.5       &   2  &  -0.506  &  0.221  	 \\
    &  1.5   &  1.5    & 2.5   &  1.5       &   2  &  0.506   &  0.221   \\
 &  1.5   &  1.5    & 1.5   &  2.5       &   2  &-0.506       &0.221     \\
\hline  
\end{tabular}
\end{center}
\caption{\textsf{$3d$ $^2D$  term:  Mg II collisional rates for the multi-term case.}}
\label{table_3d}
\end{table} 
\begin{table}
\begin{center}
\begin{tabular}{l c c c c c c r}
\hline
\hline
Term &$J$ & $J' $ &$J''$ & $J''' $ &$k$&  $a^k$ & $\lambda^k$  \\
\hline
 $4d$ $^2D$  &    2.5   &  1.5    & 2.5   &  1.5       &   2  &   14.25   & 0.386    	 \\
&  2.5   &  1.5    & 2.5   &  1.5       &   4  &   6.84   &    0.383	 \\
&  1.5   &  2.5    & 1.5   &  2.5       &   2  &     14.25   & 0.386    	 \\
&  1.5   &  2.5    & 1.5   &  2.5       &  4  &     6.84   &    0.383             \\
&  1.5   &  1.5    & 1.5   &  1.5       &  2  &   14.26 &  0.388        \\
&  2.5   &  2.5    & 2.5   &  2.5       &   2  &       10.75 &0.382       	 \\
&  2.5   &  2.5    & 2.5   &  2.5       &   4  &     11.43 & 0.376     	 \\
&  2.5   &  2.5    & 2.5   &  1.5       &   2  &   6.04   &    0.356	 \\
&  2.5   &  2.5    & 2.5   &  1.5       &   4  &    -1.85   &     0.329	 \\
&  2.5   &  2.5    & 1.5   &  2.5       &   2  &    -6.04  &  0.356   \\
&  2.5   &  2.5    & 1.5   &  2.5       &   4  &   1.85  &     0.329	 \\
&  2.5   &  2.5    & 1.5   &  1.5       &   0  &     10.26 & 0.366   	 \\
&  2.5   &  2.5    & 1.5   &  1.5       &   2  &     4.62 & 0.370    	 \\
&  2.5 & 1.5   & 2.5 & 2.5       &   2  &    6.04 &    0.356	 \\
&  2.5 & 1.5   & 2.5 & 2.5       &   4  &   -1.85    &   0.329 	 \\
   &  2.5 & 1.5   & 1.5 & 2.5       &   2  & {6.05}   & {0.357}   	 \\
   &  2.5 & 1.5   & 1.5 & 2.5       &   4  &   -0.65 &   0.390 	 \\
    &  2.5 & 1.5   & 1.5 & 1.5       &   2  &  0.008  &   0.363   	 \\
    &  1.5 & 2.5   & 2.5 & 2.5       &   2  &   -6.04 &   0.356   	 \\
  &  1.5 & 2.5   & 2.5 & 2.5       &   4  & 1.85 &    0.329  	 \\
     &  1.5 & 2.5   & 2.5 & 1.5       &   2  &   {6.05}   & {0.357}       	 \\
     &  1.5 & 2.5   & 2.5 & 1.5       &   4  & -0.65  &  0.390  	 \\
    &  1.5 & 2.5   & 1.5 & 1.5       &   2  & - 0.008  &   0.363       	 \\
    &  1.5   &  1.5    & 2.5   &  1.5       &   2  &  0.008  &   0.363        \\
 &  1.5   &  1.5    & 1.5   &  2.5       &   2  & -0.008  &   0.363         \\
\hline  
\end{tabular}
\end{center}
\caption{\textsf{$4d$ $^2D$ term:  Mg II collisional rates for the multi-term case.}}
\label{table_4d}
\end{table}

\begin{table}
\begin{center}
\begin{tabular}{l c c c c c c r}
\hline
\hline
Term &$J$ & $J' $ &$J''$ & $J''' $ &$k$&  $a^k$ & $\lambda^k$  \\
\hline
 $4f$ $^2F$  &    3.5   &  2.5    & 3.5   &  2.5       &   2  &  9.36     & 0.331   	 \\
&  3.5   &  2.5    & 3.5   &  2.5       &   4  &      7.22  &    	0.358	 \\
&  3.5   &  2.5    & 3.5   &  2.5       &   6  &      -3.68    &    	0.398	 \\
&  2.5   &  3.5    & 2.5   &  3.5       &   2  &   9.36    &    	0.331   	 \\
&  2.5   &  3.5    & 2.5   &  3.5       &  4  &      7.22    &    	0.358            \\
&  2.5   &  3.5    & 2.5   &  3.5       &  6  &        -3.68     &    	0.398            \\
&  2.5   &  2.5    & 2.5   &  2.5       &  2  &        5.80 & 0.401    	    \\
&  2.5   &  2.5    & 2.5   &  2.5       &  4  &        5.55 & 0.373    	    \\
&  3.5   &  3.5    & 3.5   &  3.5       &   2  &          9.99 & 0.367    	      	 \\
&  3.5   &  3.5    & 3.5   &  3.5       &   4  &        4.65 & 0.390    	  	 \\
&  3.5   &  3.5    & 3.5   &  3.5       &   6  &          2.90 & 0.383      	  	 \\
&  3.5   &  3.5    & 3.5   &  2.5       &   2  &   -2.17    &   0.423 		 \\
&  3.5   &  3.5    & 3.5   &  2.5       &   4  &    2.10    &    0.435	 \\
&  3.5   &  3.5    & 3.5   &  2.5       &   6  &    -12.51     &   0.382 	 \\
&  3.5   &  3.5    & 2.5   &  3.5       &   2  &    2.17     &    0.423  	 \\
&  3.5   &  3.5    & 2.5   &  3.5       &   4  &   -2.10    &     0.435		 \\
&  3.5   &  3.5    & 2.5   &  3.5       &   6  &    12.51   &    0.382 			 \\
&  2.5   &  2.5    & 3.5   &  3.5       &   0  &          12.02 &0.376    	  	 \\
&  2.5   &  2.5    & 3.5   &  3.5       &   2  &         1.34 &0.379    	    	 \\
&2.5   &  2.5    & 3.5   &  3.5       &   4  &         0.44 & 0.367   	    	 \\
&  3.5 & 2.5   & 3.5 & 3.5       &   2  &    -2.17    &    	 0.423 	 \\
&  3.5 & 2.5   & 3.5 & 3.5       &   4  &   2.10     &    	 0.435 	 \\
&  3.5 & 2.5   & 3.5 & 3.5       &   6  &   -12.51     &    	 0.382 		 \\
   &  3.5 & 2.5   & 2.5 & 3.5       &   2  &   0.234   &   0.469 	   	 \\
   &  3.5 & 2.5   & 2.5 & 3.5       &   4  &     0.889   &    0.298	 	 \\
      &  3.5 & 2.5   & 2.5 & 3.5       &   6  &    - 3.61    &    0.368	 	 \\
    &  3.5 & 2.5   & 2.5 & 2.5       &   2  &     2.75  &    	 0.388  	 \\
        &  3.5 & 2.5   & 2.5 & 2.5       &   4  &    -0.997    &    0.266	   	 \\
    &  2.5 & 3.5   & 3.5 & 3.5       &   2  &     2.17   &    0.423 	   	 \\
  &  2.5 & 3.5   & 3.5 & 3.5       &   4  &   -2.10   &  0.435   	  	 \\
    &  2.5 & 3.5   & 3.5 & 3.5       &   6  & 12.51   &    	0.382 	  	 \\
     &  2.5 & 3.5   & 3.5 & 2.5       &   2  &   0.234    &     0.469 	  	     	 \\
     &  2.5 & 3.5   & 3.5 & 2.5       &   4  &    0.889  &    0.298	  	 \\
         &  2.5 & 3.5   & 3.5 & 2.5       &   6  &    -3.61  &    	 0.368	   	 \\
    &  2.5 & 3.5   & 2.5 & 2.5       &   2  &   -2.75  &    0.388	       	 \\
        &  2.5 & 3.5   & 2.5 & 2.5       &   4  &  0.997  &    0.266	       	 \\
    &  2.5   &  2.5    & 3.5   &  2.5       &   2  &  2.75     &  0.388  	       \\
       &  2.5   &  2.5    & 3.5   &  2.5       &   4  &       -0.997 &   0.266 	       \\
 &  2.5   &  2.5    & 2.5   &  3.5       &   2  &   -2.75   &    	0.388        \\
  &  2.5   &  2.5    & 2.5   &  3.5       &   4  &   0.997   &    	0.266        \\
\hline  
\end{tabular}
\end{center}
\caption{\textsf{$4f$ $^2F$  term:  Mg II collisional rates for the multi-term case.}}
\label{table_4f}
\end{table} 

\end{appendix}

\end{document}